\documentclass[useAMS,usenatbib,usegraphicx]{mn2e}

\usepackage{psfig}

\newcommand{\HI}{\mbox{H\,{\sc i}}}
\newcommand{\Lya}{Lyman $\alpha$}
\newcommand{\Lyb}{Lyman $\beta$}

\newcommand{\HeII}{\mbox{He\,{\sc ii}}}
\newcommand{\CIV}{\mbox{C\,{\sc iv}}}

\newcommand{\kms}{km~s$^{-1}$}

\def\pa{{\parallel}}
\def\pe{{\perp}}
\def\los{line of sight}
\def\loss{lines of sight}

\title[Tomography of the intergalactic medium]{Tomography of the
  intergalactic medium with Ly$\alpha$ forests in close QSO pairs
  \thanks{Based on observations collected at the
  European Southern Observatory Very Large Telescope, Cerro Paranal,
  Chile -- Programs  65.O-0299(A), 68.A-0216(A), 69.A-0204(A),
  69.A-0586(A), 70.A-0031(A), 166.A-0106(A) }} 
\author[V. D'Odorico et al.]{V. D'Odorico$^{1}$\thanks{E-mail:
dodorico@oats.inaf.it}, M. Viel$^{2,1}$, F. Saitta$^3$,
  S. Cristiani$^{1}$, S. Bianchi$^4$, B. Boyle$^{5,6}$, 
\newauthor S. Lopez$^7$, J. Maza$^7$, P. Outram$^8$ \\
$^1$INAF -- Osservatorio Astronomico di Trieste, via G.B. Tiepolo 11,
Trieste, I-34131, Italy\\
$^2$Institute of Astronomy, Madingley Road, Cambridge CB3 0HA \\
$^3$Dipartimento di Astronomia, Universit\`a degli Studi di Trieste, via
  G.B. Tiepolo, 11, Trieste, I-34131, Italy \\ 
$^4$INAF - Istituto di Radioastronomia, Sezione di Firenze, Largo E. Fermi 5,
Firenze, I-50125, Italy \\
$^5$ Australia Telescope National Facility, PO Box 76, Epping NSW
  1710, Australia\\
$^6$ Anglo-Australian Observatory, PO Box 296, Epping,
  NSW 2121, Australia \\
$^7$Departamento de Astronom\'\i a, Universidad de Chile, Casilla 36-D,
  Santiago, Chile \\
$^8$Department of Physics, University of Durham, Science Laboratories,
  South Road, Durham, DH1 3LE }
 
\begin{document}

\date{Accepted 2006 August 14. Received 2006 August 9; in original
  form 2006 March 17}

\pagerange{\pageref{firstpage}--\pageref{lastpage}} \pubyear{2006}

\maketitle

\label{firstpage}

\begin{abstract}
We study the 3-dimensional distribution of non virialised matter at
$z\sim2$ using high resolution spectra of QSO pairs and simulated
spectra drawn from cosmological hydrodynamical simulations.
We have collected the largest sample of QSO pairs ever observed with
UVES at the ESO-VLT, with angular separations between $\sim 1$ and 14
arcmin.   
The observed correlation functions of the transmitted flux in the \HI\
\Lya\ forest along and transverse to the lines of sight are in
good agreement implying that the distortions in redshift space due to 
peculiar velocities are small. The clustering signal is significant
up to velocity separations of $\sim 200$ \kms, or about $3\ h^{-1}$
comoving Mpc.  The regions at lower overdensity ($\rho/\bar{\rho} \la
6.5$) are still clustered but on smaller scales ($\Delta v \la 100$ \kms). 
The observed and simulated correlation
functions are compatible at the $3 \sigma$ level. A better concordance
is obtained when only the low overdensity regions are selected for
the analysis or when the effective optical depth of the simulated
spectra is increased artificially, suggesting a deficiency of strong
lines in the simulated spectra. 
We found that also a lower value of the power-law index of the
temperature-density relation for the \Lya\ forest gas improves the
agreement between observed and simulated results. 
If confirmed, this would be consistent with other observations favouring 
a late  \HeII\ reionization epoch (at $z\sim 3$).      
We remark the detection of a significant clustering signal in the
cross correlation coefficient at a transverse velocity separation
$\Delta v_{\pe} \sim 500$ \kms\ whose origin needs further investigation. 
\end{abstract}


\section{Introduction}

The study of the evolution of cosmic structures is one of the major
topic of present-day cosmology.  




It is now more than a decade that thanks to semi-analytical and
hydrodynamical simulation results \citep[e.g.,][]{cen94,zhang95,
hernquist96,miralda96,bi97,dave97,zhang97,theuns98a,machacek00}  the majority 
of the absorption features observed in \Lya\ forests of high redshift
QSO spectra 
is identified with the fluctuations of the intermediate and low density
intergalactic medium (IGM), arising naturally in the hierarchical
process of structure formation. 
The physics of this highly ionized gas is simple, governed mainly by the
Hubble expansion and the gravitational instability  \citep{rauch05}.
If photoionization equilibrium is assumed, a mean relation between
temperature and overdensity is obtained \citep{huignedin}:  

\begin{equation}
T = T_0\ (\rho/  \bar{\rho})^{\gamma -1},
\end{equation}

\noindent 
where $T_0$ and $\gamma$ depend on the ionization history of the
Universe (for an early reionization,  $T_0=10^4$ K  and $\gamma=1.6$ are generally 
adopted) . The combination of photoionization equilibrium and 
eq.~(1) leads to a power law relation between measured \HI\ \Lya\ optical 
depth and gas overdensity \citep[see e.g.][ for a review]{weinberg99},
which, adopting the standard values for the parameters, can
be written as: 

\begin{equation} 
\tau_{\rm H\,I} \simeq 1.3 \times 10^{-3} (1+z)^{9/2}\,(\rho/
\bar{\rho})^{3/2}. 
\end {equation}

A critical test of the nature of the \Lya\ absorbers as proposed 
by simulations, comes from the determination of their sizes 
and spatial distribution.  
Correlations in the  \Lya\ forest were detected with a 4--5$\
\sigma$ confidence by various authors at typical scales $\Delta v \la
350$ \kms\ observing at high resolution individual lines of
sight (\citealt{lu96} at $z\sim 3.7$; \citealt{cristiani97} at
$z\sim3$; \citealt{kim01} at $z\sim2$). This velocity range
corresponds to scales $\la 2.5\ h^{-1}$ Mpc (considering peculiar
velocities negligible). In this
observational approach, however, the 3-dimensional information is
convoluted with distortions in the redshift space, due to peculiar
motions and thermal broadening.  

Multiple \loss\ offer an invaluable alternative to address the spatial
distribution of the absorbers, enabling a more direct interpretation
of the observed correlations. 
The interesting range of separations lies about the Jeans scale of 
the photoionized IGM ($\sim 1$ arcmin or $\sim 1.4\ h^{-1}$
comoving Mpc  at $z\sim2$). At this scale there should 
be a transition from a smooth gas distribution, which produces nearly 
identical absorption features in neighbouring \loss, to a correlated 
density distribution, where the correlation strength decreases with
increasing separation of the \loss\ \citep[e.g., ][]{viel2002}. 
Indeed, the spectra of multiple images of lensed quasars with
separations of the order of a few arcsec 
\citep{smette92,smette95,impey96,rauch01} show nearly identical \Lya\
forests, implying that the absorbing objects have large sizes ($> 50\
h^{-1}$~kpc).  
For pairs with larger separation the correlation between the
absorption features becomes weaker. Studies based on the statistics of   
coincident and anti-coincident absorption lines provided evidence for  
dimensions of a few hundred kpc
\citep{bechtold94,dinshaw94,dinshaw95,dinshaw97,fang96,crottsfang98,
vale98,petitjean98,lopez00,young01,becker04}. 
The obtained large sizes were conclusive to exclude models of the
\Lya\ forest as a population  
of pressure confined small clouds \citep[e. g.,][]{sargent80} or clouds in 
dark matter mini haloes \citep[e. g.,][]{miralda93}.  

The new theoretical scenario and the production of exceptional quality
QSO spectra by high resolution spectrographs at 10-m class
telescopes, started the use of the \Lya\ forest as a cosmological
probe. From the study of absorption spectra along single \loss\ to
distant QSOs and the implementation of various algorithms to convert
the observed flux into density it was possible, for example, to
determine the shape and amplitude of the power spectrum of the spatial
distribution of dark matter  
\citep*{croft98,croft99,croft02,nusserhaehnelt99,nusserhaehnelt00,viel2004a}
or, more straightforwardly, of the transmitted flux
\citep{mcdonald00,kim04}.    

The final goal of this paper is to investigate the distribution
properties of matter in the IGM applying the modern interpretation of 
the \Lya\ forest. In particular, we computed the correlation
properties of the transmitted flux along and across the \loss\ for a new
exceptional sample of high resolution spectra of close QSO pairs and
groups at $z_{\rm  em} \sim 2$.    
Previous works following a similar approach
\citep{rollinde03,coppolani06} used sets of low resolution spectra of
QSO pairs with the main aim of deriving constraints on
$\Omega_{0\Lambda}$ from the application of the Alcock-Paczy\'nski
test \citep[][ see also Section~6]{ap}.   
However, about $13(\Delta\theta/(1$ arcmin)$)^2$ QSO pairs with
separation $<\Delta\theta$ are needed to determine
$\Omega_{0\Lambda}$ with a precision better than 10 per cent
\citep{mcdonald03}. So, we will not attempt to determine
$\Omega_{0\Lambda}$ in this study, although our plan is to gather a
larger sample of QSO pairs in a few years.  

The present investigation is completed by the comparison with a set of
mock spectra drawn from an hydrodynamical simulation, reproducing 
several different realisations of our sample of QSO associations. We
analysed how variations of the mean flux (corresponding to variations
of the intensity of the ionizing background) and of the
temperature-density relation of the gas affect the correlation functions. 

In this preliminary study, we do not take advantage of the presence,
in our sample, of two groups of QSOs (one triplet and one sestet) to
detect structures extending on large scales.  
Since new observing time have been allocated to our project at
UVES-VLT, a forthcoming  paper will be devoted to the detailed
analysis of coincidences and anticoincidences both of \Lya\ and metal
absorptions among multiple \loss, on improved
signal-to-noise ratio (S/N) spectra.

The paper structure is as follows: in Section~2 we describe the
observed data sample, the reduction procedure and the simulated
spectra. Sections~3, 4 and 5 are devoted to the computation of the
auto and cross correlation functions of both the observed and
simulated spectra and to their comparison. 
We discuss some interesting aspects of our results and we explore the
parameter space of the simulations in Section~6. 
The conclusions are drawn in Section~7.  
   
Throughout this paper we adopt $\Omega_{0\rm{m}} = 0.3,\
  \Omega_{0\Lambda}=0.7$, and  $h=H_0/72$ km s$^{-1}$ Mpc$^{-1}$.

\begin{table}
  \centering
    \caption{Characteristics of the observed QSO spectra}
    \label{obs:qso}
    \begin{tabular}{@{}rcccc@{}}
      \hline 
    Object & $z$ & M$_{\rm B}$ & \Lya\ & S/N  \\ 
& & & range & per pixel \\
\hline 
Pair A\ \ \ \ \ \ \ \ \  PA1 & 2.645 & 19.11 & 2.094--2.585 & 8--12 \\
\ \ \ \ \ \ \ \ \ \ \ \ \ \ PA2 & 2.610 & 19.84 & 2.094--2.550 & 3.5--6.5 \\
Triplet\ \ \ \ \ \ \ \ \ \ \ T1 &  2.041 & 18.20 & 1.633--1.991 & 3--10 \\
       \ \ \ \ \ \ \ \ \ \ \ T2 &  2.05  & 18.30 & 1.592--1.999 & 3.5--8.5 \\
       \ \ \ \ \ \ \ \ \ \ \ T3 &  2.053 & 18.10 & 1.665-2.002 & 2.5--6 \\
Sestet\ \ \ \ \ \ \ \ \ \ \ \ S1 &  1.907 & 19.66 & 1.665--1.859 & 2--5.5 \\
      \ \ \ \ \ \ \ \ \ \ \ \ S2 &  2.387 & 19.53 & 1.858--2.331 & 3--7.5\\
      \ \ \ \ \ \ \ \ \ \ \ \ S3 &  2.102 & 19.31 & 1.633--2.051 & 4--10 \\
      \ \ \ \ \ \ \ \ \ \ \ \ S4 &  1.849 & 19.59 & 1.575--1.802 & 2--8 \\
      \ \ \ \ \ \ \ \ \ \ \ \ S5 &  2.121 & 18.85 & 1.633--2.069 & 3.5--8\\
      \ \ \ \ \ \ \ \ \ \ \ \ S6 &  2.068 & 20.19 & 1.592--2.017 & 3--8.5 \\
Pair U\ \ \ \ \ UM680  & 2.144 & 18.60 & 1.653--2.092 & 6.5--17 \\
      \ \ \ \ \ UM681  & 2.122 & 19.10 & 1.634--2.070 & 7--17 \\
Pair Q Q2343+12 & 2.549 &17.00 & 1.994--2.490 & 13--23 \\
       Q2344+12 & 2.773 & 17.50 & 2.183--2.711 & 12--18 \\
\hline \\
\end{tabular}
\end{table}

\section{Data sample} 

\subsection{The observed spectra}

\begin{figure*}
\includegraphics[height=8truecm,width=8.5truecm]{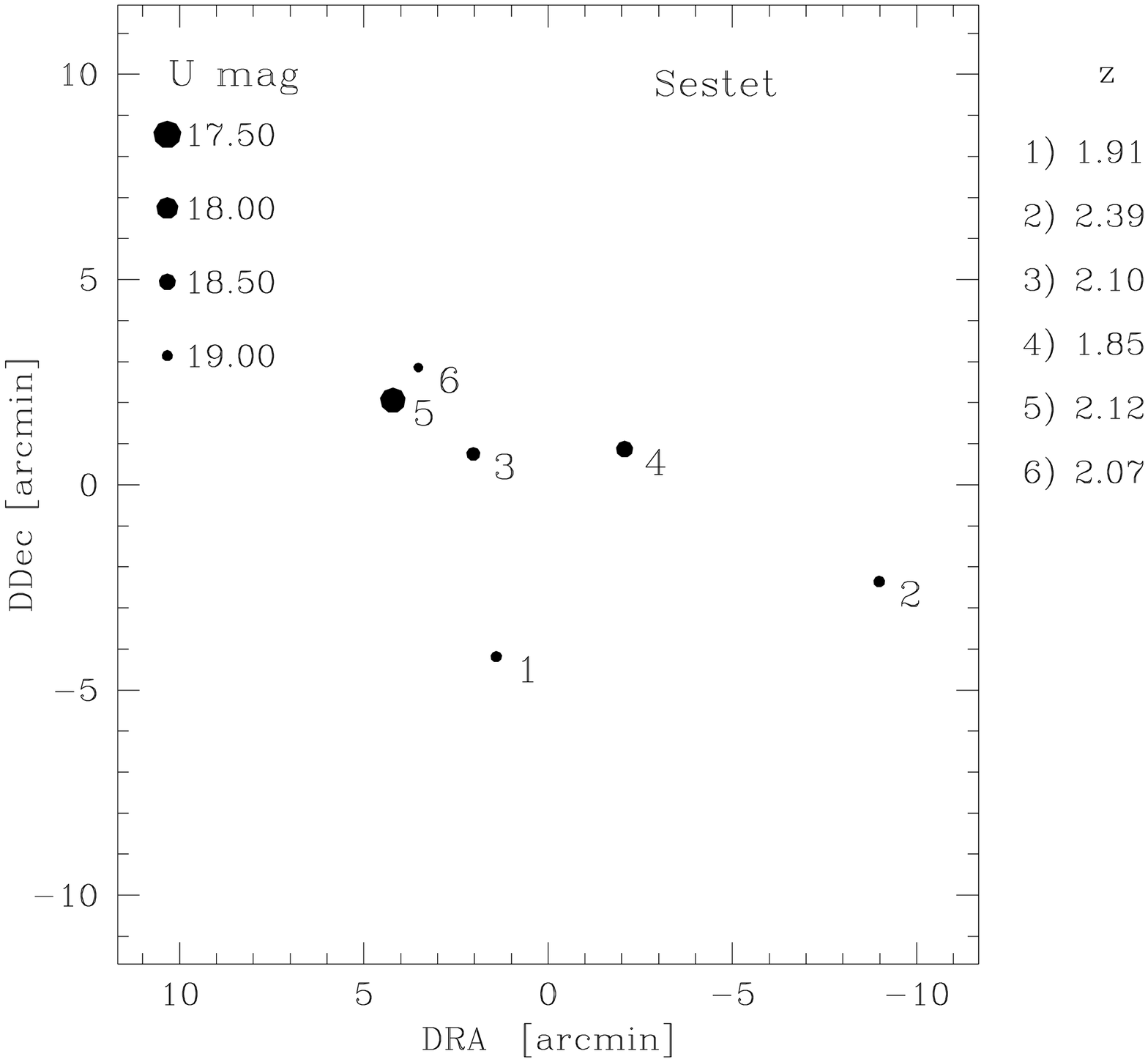}
\includegraphics[height=8truecm,width=8.5truecm]{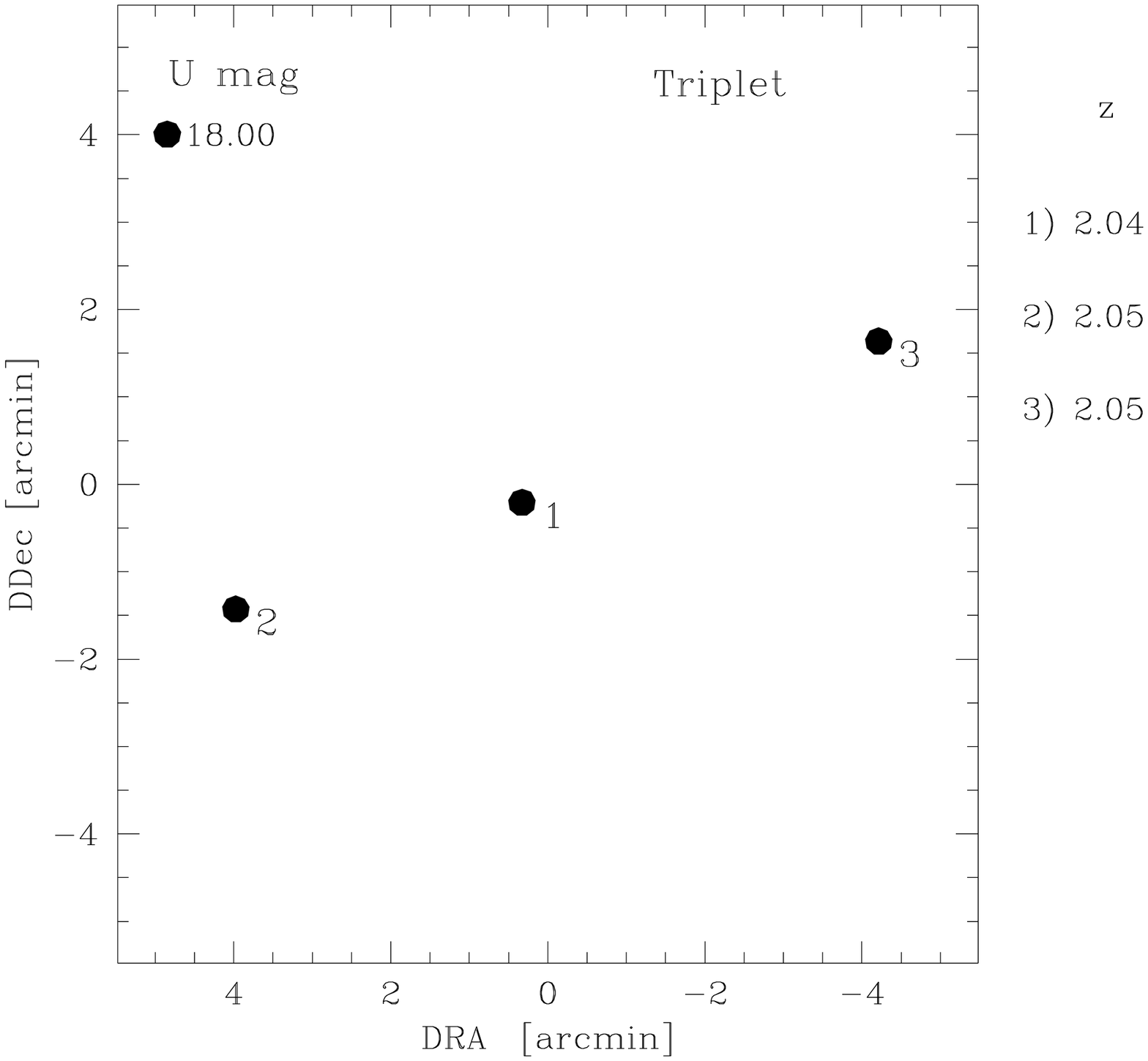}
  \caption{Relative positions, U magnitudes and redshifts of the QSOs
  composing the sestet (left plot) and the triplet (right plot) in our
  sample}\label{fig:sestet} 
\end{figure*}

The exploitation of the potential offered by multiple QSO \loss\ has 
been limited by the dearth of suitable groups of QSOs close and bright 
enough to permit high resolution spectroscopy. 
Two major breakthroughs have dramatically improved this situation:
\par\noindent
- the 2dF QSO Redshift Survey \citep[][ www.2dfquasar.org]{croom04},
whose complete spectroscopic catalogue contains more than $\sim
23000$ QSOs in a single homogeneous data base, which is approximately
50 times more than the previous largest QSO survey to a similar depth
($B<21$); 

\par\noindent
- the UVES spectrograph \citep{dekker} at the Kueyen unit of the ESO
VLT (Cerro Paranal, Chile) which has a remarkable efficiency
especially in the extreme UV (close to the atmospheric cut-off).
In this way, relatively low-redshift ($z\la 2.5$)
QSOs become accessible to high resolution observations of the \Lya\
forest down to faint apparent magnitudes and their surface density
becomes high enough to provide several \loss\ for a tomography of the
IGM. 

We searched the 2dF QSO database for the best groups with apparent
magnitude $B\le 20$ and $z>1.8$. A triplet found in the Cal\'an-Tololo 
QSO survey \citep{maza93,maza95} was added to the sample. 
A great observational effort was carried out to collect UVES
spectra of the selected QSOs, which all have magnitudes fainter than
$B \sim 19$, with the exception of the triplet for which $B\sim 18$.   
ESO allocated 51 hours of observation to our project up to now, which
allowed us to obtain acceptable S/N spectra of
one pair (from now on called Pair A, with angular separation of $\sim
1.3$ \arcmin), one sestet and the triplet of QSOs (see
Fig.~\ref{fig:sestet}).   

Reduction of the QSO spectra was conducted with the pipeline
of the instrument \citep[version 2.1, ][]{ball00} provided by ESO in 
the context of the data reduction package MIDAS.  
In most cases, we could not apply the standard procedure due to the
faintness of the spectra in the blue region and had to pre-filter the
cosmic rays to make the optimal extraction work properly.  
Single extracted spectra were summed and rebinned and wavelengths were
corrected to the vacuum-heliocentric reference frame.  
The final spectra have resolution $R\sim 40000$  in the \Lya\ and
\CIV\ forest, while the S/N per pixel varies on
average between $S/N \sim 3$  and 10 in the \Lya\ forest and between 4
and 15 in the \CIV\ forest (see Table~\ref{obs:qso} for details).

Continuum determination, in particular in the \Lya\ forest region, is a 
very delicate step in the process of spectra reduction. 
Tentative procedures realised up to now to objectively determine the
continuum position through automatic algorithms do not give
satisfactory results. 
We adopted a manual subjective method based on the selection of the
regions free from clear absorption that are successively fitted with a
spline polynomial of 3rd degree. 

The limitations introduced by the uncertainty in the true continuum
level and shape should be less important in the computation of the
cross correlation function than in the case of single \los\ 
analysis. This because the undulations are 
uncorrelated between adjacent \loss, in particular, if the two QSOs do
not have too similar redshifts \citep[see][ for a discussion in the case
  of power spectra]{viel2002}. 


We added to the sample two more pairs from our archive: UM680/UM681
(Pair U, separated by  $\sim 1$ arcmin) and Q2344+1228/Q2343+1232 (Pair Q, 
separated by $\sim 5.57$ arcmin). They were observed with UVES at the same
resolution as the new data, and reduced following the same
procedure. More details on their properties are given in the paper by   
\citet{vale02}. 

\begin{figure}
\includegraphics[height=8truecm,width=8.5truecm]{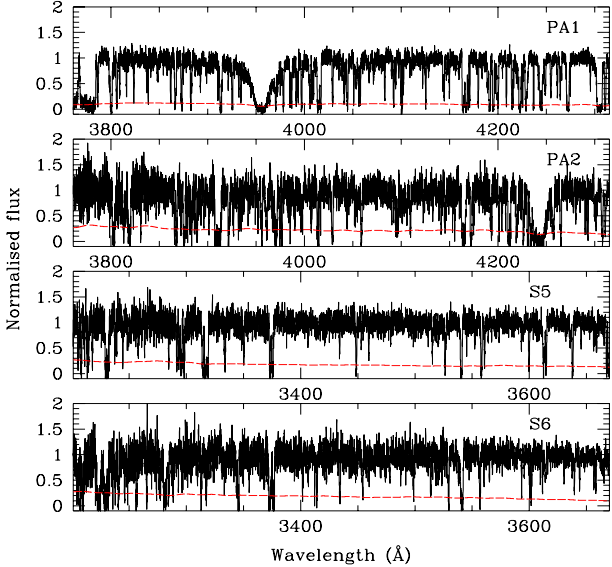}
  \caption{\Lya-forests of two of the closest pairs in our
  sample. Pair A (upper two panels) separated by 1.3\arcmin\ and the
  pair at 1.06\arcmin\ formed by the QSOs S5 and S6 of the sestet
  (lower two panels)}\label{fig:spectra} 
\end{figure}

The list of the QSOs and the main characteristics of the spectra are
reported in Table~\ref{obs:qso}. The \Lya\ forests of two of the
closest pairs in our sample are shown in Fig.~\ref{fig:spectra} 

The total sample is formed by 21 pairs uniformly distributed between
angular separations of $\sim 1$ and 14 arcmin, corresponding to
comoving spatial separations between $\sim 1.4$ and $21.6\
h^{-1}$ Mpc. 
The median redhift of the \Lya\ forest is $z \sim 1.8$. 
This is the largest sample of high resolution spectra of QSO pairs
ever collected, unique both for the number density -- we have 6 QSOs in
a region of $\sim 0.04$ square degrees -- and the variety of \los\
separations investigated.    

\begin{table}
  \centering
    \caption{List of the considered QSOs from the ESO
    Large Program (see text)}
    \label{lp:qso}
    \begin{tabular}{@{}lccl@{}}
      \hline 
    Object & Redshift & M$_{\rm B}$ & \Lya\ range \\ 
\hline 
HE 1341-1020 & 2.134 & 17.1  & 1.64431--2.08216 \\
Q 0122-380   & 2.189 & 16.7  & 1.69072--2.13625 \\
PKS 1448-232 & 2.220 & 16.96 & 1.71688--2.16674 \\
HE 0001-2340 & 2.280 & 16.7  & 1.76750--2.22575 \\
Q 0109-3518  & 2.406 & 16.44 & 1.87381--2.34966 \\
HE 2217-2818 & 2.406 & 16.0  & 1.87381--2.34966 \\
Q 0329-385   & 2.423 & 16.92 & 1.88816--2.36638 \\
HE 1158-1843 & 2.453 & 16.93 & 1.91347--2.39589 \\
\hline 
\end{tabular}
\end{table}

As a reference sample (see Sections~4 and 5.1), we considered 8
UVES QSO spectra obtained in the framework of the ESO 
Large Program `The Cosmic Evolution of the IGM'
\citep[][ LP sample]{bergeron04}. The QSOs were selected in order to
match as much as possible the redshift range of our sample 
(see Table~\ref{lp:qso}).  The spectra have resolution $R\sim 45000$
and S/N ratio $\sim 50$ in the \Lya\ forest region. They were
normalised to the continuum and prepared for the analysis following
the same procedure adopted for the spectra of the pair sample. 

\subsection{The simulated spectra}

In order to both assess the nature of the \Lya\ forest inferred from 
simulations and to constrain the cosmological scenario of the same 
simulations, we compared the results obtained for our sample of 
observed QSO spectra with analogous results for a sample of mock 
\Lya\ forests. 

We used simulations run with the parallel hydrodynamical (TreeSPH)
code {\small {GADGET-2}} \citep{springel2001,springel2005}.  The
simulations were performed with periodic boundary conditions with an
equal number of dark matter and gas particles and used the
conservative `entropy-formulation' of SPH proposed by
\citet{springel2002}. 
Radiative cooling and heating processes were followed for a primordial
mix of hydrogen and helium. We assumed a mean UV background
produced by quasars and galaxies as given by \citet{hm96}
with helium heating rates multiplied by a factor 3.3 in order to fit
observational constraints on the temperature evolution of the IGM.  
More details can be found in \citet{viel2004a}. 

The cosmological model corresponds to a `fiducial' $\Lambda$CDM
Universe with parameters $\Omega_{0\rm{m}}=0.26,\
\Omega_{0\Lambda}=0.74,\ \Omega_{0\rm{b}}=0.0463$ and $H_0 = 72$ km
s$^{-1}$ Mpc$^{-1}$ (the B2 series of \citealt{viel2004a}). 
We have used $2\times 400^3$ dark matter and
gas particles in a $120\ h^{-1}$ comoving Mpc box. 
The gravitational softening was set to 5 $h^{-1}$ kpc in
comoving units for all particles.
We note that the parameters chosen here, including the thermal
history of the IGM, are in perfect agreement with observational
constraints including recent results on the CMB and other results
obtained by the \Lya\ forest community
\citep[e.g. ][]{spergel,viel2004a,seljak}. 

The $z=1.8$ output of the simulated box was pierced to obtain 50
triplets of \loss\ carefully reproducing the observed triplet mutual
separations.   
The same was done for 50 sestets of \loss\ reproducing the observed
sestet and 50 pairs of \loss\ at the same angular separation as Pair U. 
Fifty different realisations of  Pair A spectra and 50 of Pair Q were
obtained from the output box at redshift $z=2.4$.    

The velocity extent of the spectra is of 10882.3 \kms\ for the sestet, 
the triplet and Pair U, and of  11683.9 \kms\  for Pair A and Pair Q.  
The pixel velocity size is fixed for each spectrum and slightly
increases with the redshift of the box, being ${\rm d}v \simeq 2.66$,
$2.85$ \kms\ for $z = 1.8$, $2.4$ respectively.  
These values are comparable with the average pixel size of the observed
spectra, ${\rm d}v_{\rm obs} \simeq 2.64$ \kms.     

The mean \HI\ optical depth in the simulation box is determined by the 
UV background at the considered redshift. The corresponding observed 
quantity is the so called effective optical depth, $\tau_{\rm eff}$,
which is obtained from the average flux measured in QSO absorption
spectra: $\bar{f} = \exp(-\tau_{\rm eff})$. 
The $\tau_{\rm eff}$ of the simulated spectra can be rescaled to a 
given value simply by multiplying the optical depth of every pixel by
a constant factor \citep[e.g.][]{theuns98b,bolton05}.
So for the
spectra reproducing our sample of QSOs we have adopted the average
values measured in the observed spectra: $\tau_{\rm eff} =  0.10$ for
the triplet and Pair U (average \Lya\ redshift $z_{\rm Lya} \simeq
1.83$), $0.12$ for the sestet ($z_{\rm Lya}\simeq 1.83$), and
$0.21$ for  Pair A and Pair Q ($z_{\rm Lya}\simeq 2.34$). 
The chosen values are within 1 per cent of the original values of the
simulated spectra.  
In the computation of the correlation functions, we have used the
average fluxes obtained as the mean of the flux values along the \los\ 
as for the observed spectra. 
We checked that the difference in the resulting correlation functions
taking the true values for the average flux is less than 5 per cent.   

Finally, we added to the simulated spectra a Gaussian noise in order
to reproduce the observed average S/N (per pixel):
$S/N=5$  for the triplet, the sestet and Pair A; $S/N=9$ for Pair U
and $S/N=15$ for Pair Q. 


\section{Correlation properties of the IGM: general procedure}

On the basis of the interpretation of the \Lya\ forest as due to a
continuous density field with a univocal correspondence between
density and transmitted flux, we computed the correlation properties 
of the transmitted flux in QSO \loss\ and regarded them as indicators of 
the correlation properties of matter in the IGM.

We selected in each normalised spectrum the region between the 
\Lyb\ emission (or the shortest observed wavelength, when the \Lyb\
was not included in the spectrum) and 5000 km s$^{-1}$  
from the \Lya\ emission (to avoid proximity effect due to the
QSO). Absorption lines due to ions of elements heavier than hydrogen
`contaminate' the \Lya\ forest and can give spurious contribution to
the clustering signal \citep[see][ for a discussion in the case of
  single \loss]{kim04}.   
We flagged and removed the spectral regions where metal lines and
\Lya\ absorptions of damped and sub-damped systems occurred inside the
\Lya\ forest. 

Given the normalised transmitted flux, $f$, as a function of
the velocity $v_{\pa}$ along the \los\ and the angular position
$\theta$ on the sky, we define $\delta_f = (f - \bar{f})$,  
%
%
%
%
where the average flux, $\bar{f}$, is computed for every spectrum as
the mean of the transmitted flux over all the considered 
pixels in that spectrum. 
We neglected the redshift evolution of the average transmitted
flux in the \Lya\ forest of the individual spectra, which translates
into the redshift evolution of the mean \HI\ opacity of the Universe
\citep{kim02,schaye03,viel2004a}, because we verified that its effect 
on the correlation function is negligible.  

\begin{table}
  \centering
    \caption{Meaning of the symbols used in the following figures}
    \label{symbols}
    \begin{tabular}{@{}llll@{}}
      \hline 
   & LP sample & pair sample  & simulations \\ 
\hline 
Auto correl. & cross & solid triangle & empty triangle \\
Cross correl. &  & solid square & empty square \\
Cross corr. coeff. & & solid dot & empty dot \\
\hline 
\end{tabular}
\end{table}

Table~\ref{symbols} provides details about the symbol convention used
to plot our results.


\section{The flux auto correlation function}


The unnormalised auto correlation function of the flux along the \los\
is defined as: 

\begin{equation}
\xi^f_{\pa}(\Delta v_{\pa}) = \langle \delta_f(v_{\pa},\theta)
\delta_f(v_{\pa}+\Delta v_{\pa},\theta)\rangle,  
\end{equation}

\par\noindent
following previous studies on the same subject
\citep[e.g.][]{mcdonald00,rollinde03,becker04}.  
The auto correlation function for our sample of QSO spectra was
obtained by averaging over all the pixels of all the QSOs. 
The results were binned in 50 \kms\ velocity bins.  
The 1~$\sigma$ error on this measure is estimated extracting 50
samples of 15 objects from our sample with a 
bootstrap method, computing the auto correlation function for each
sample and determining the standard deviation of the distribution.

The obtained correlation function is in very good agreement with the
analogous determination by \citet{rollinde03} and in qualitative
agreement with the \Lya\ lines two point correlation function at
similar redshifts \citep{kim01}. 

The same procedure was adopted to compute $\xi^f_{\pa}$ for the QSO
spectra of the control sample taken from the UVES LP. 
These spectra have a much larger S/N ($\sim 50$
in the \Lya\ forest region) than our spectra, they have been selected 
to be free from damped \Lya\ absorptions and they do not belong to
known close QSO groups. 

\begin{figure}
\includegraphics[height=8truecm,width=8truecm]{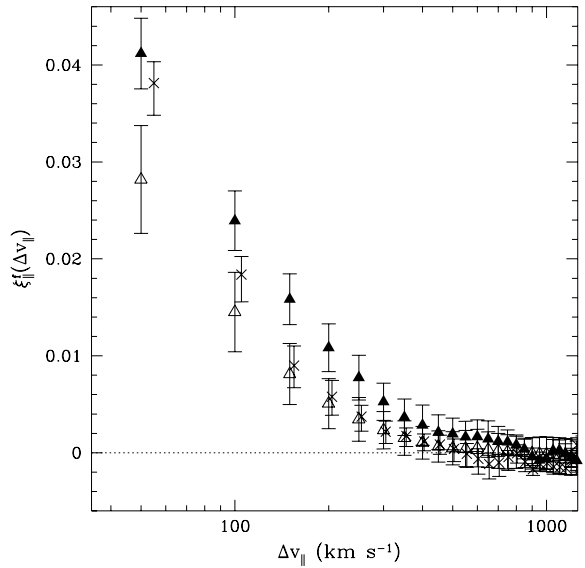}
  \caption{Comparison of the auto correlation function for our sample
  of observed QSO spectra (solid triangles), for the LP sample of
  observed QSO spectra (crosses, shifted in velocity for clarity) and
  for the simulated sample of spectra (empty triangles) as a function of velocity
  separation along the \los.  Error bars on $\xi^f_{\pa}$ are highly
  correlated and they are 
  plotted only to give an idea of their extent.}\label{fig:comp_autov} 
\end{figure}

The auto correlation function for the simulated spectra was
computed as the arithmetic mean of the correlation functions obtained
for 50 realisations of the observed sample and the error is the
corresponding standard deviation.    
It is important to recognise that the computed error bars both for 
the observed and simulated $\xi^f_{\pa}$ are strongly correlated. This is
due to the fact that every pixel contributes to the correlation
function in several velocity bins. 

In Fig.~\ref{fig:comp_autov} we show the comparison of the three 
auto correlation functions in velocity space, error bars are included
just to give a qualitative idea of their size.    

The agreement at the $3\ \sigma$ level is good.   
However, we would like to point out two kinds of discrepancies at the
$1\ \sigma$ level which, in our opinion, are the signatures of two
interesting effects. 
\par\noindent
1. Both observed correlation functions have an amplitude in the
first bin (centred at 50 \kms) which is larger than the simulated
one.  We ascribe this to a scantiness of strong lines ($\log
N($HI$)\ga 15.5$) in the simulated spectra \citep[see
  e.g. ][]{theuns02} due to the limited extent of simulations in
comoving space, which implies that large over densities (giving rise
to strong \Lya\ lines) are not appropriately probed. 
We found support to this hypothesis from the analysis in Sections~4.1 and
6.1.1.  
\par\noindent
2. The correlation function computed with our sample of QSOs tends to  
show a stronger clustering than the simulated one up to large
separations, while the LP correlation function is compatible with
the mock correlation function already at the second bin. 
The main difference between the two samples is the fact that our QSOs  
were selected to belong to close pairs and associations, thus they
could trace over-clustered regions. Although we excluded the spectral
portion within 5000 \kms\ of the QSO emission, an enhanced clustering
signal could still be present. We discuss this hypothesis in
Section~4.2.   

We notice that cosmic variance could also be at the origin of these
small differences. Indeed auto correlation functions of single
observed QSO spectra show large variations among them and we are
working with relatively small samples. 
On the other hand, all the simulated spectra have been drawn from the
same box of 120 $h^{-1}$ comoving Mpc, which although notably large
for a simulation, possibly underestimates the cosmic variance
\citep[see][ for possible ways of correcting the 
simulated correlation function for errors induced by the finite box
size]{mcdonald00}.  


\subsection{Contribution of the stronger absorptions}

\begin{figure}
\includegraphics[height=8truecm,width=8truecm]{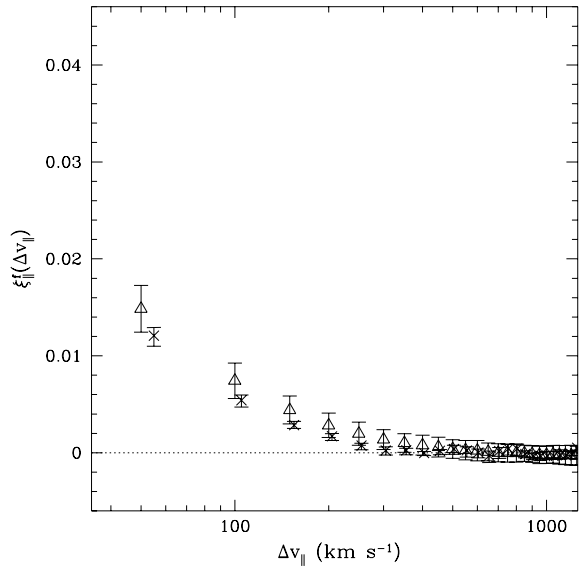} 
  \caption{Same as Fig.~\ref{fig:comp_autov} but selecting only the
  pixels with flux value $f > 0.1$ both in the observed and simulated
  spectra. This is equivalent to exclude the lines with column density
  $\log N($\HI$)\ga 13.96$.}\label{fig:comp_autov_f01} 
\end{figure}

\Lya\ lines of increasing \HI\ column density correspond to structures  
of increasing overdensity and smaller characteristic size. Going
from $\log N($\HI$) \sim 12$ to $\log N($\HI$) \ga 15$ we probe
from under-dense, modestly over-dense regions, to highly 
over-dense filaments and their intersections.  
We expect that the major contributors to the clustering signal are 
the stronger absorption lines, which trace the higher density
peaks. The aim of this section is twofold. On the one hand, by
excluding the stronger lines we want to verify the clustering
properties of the gas in the `true' IGM, far from the densest
regions. On the other hand,  we want to test the relative behaviour of
the observed and simulated auto correlation  functions without those
lines.  

Note that a careful study of the impact of strong absorption systems 
on the flux power spectrum was performed by \citet{viel2004b} and
\citet{mcdonald05}.  

To select the strong absorption lines without carrying on the time 
consuming Voigt fitting procedure of the observed and simulated
spectra, we can relate the transmitted flux values to a 
corresponding column density  assuming a typical Doppler parameter. 
We eliminated the pixels with $f < f_{\rm thres} = 0.1$ or $\tau_{\rm
  H\,I} \ga 2.3$. 
This approximately corresponds to excluding the absorption lines with
$\log N($\HI$) \ga 13.96$, if we adopt the formula for the optical 
depth at the centre of the line: 

\begin{equation}
\tau_{0,\rm{Ly}-{\alpha}} = 7.58295\times 10^{-13} \frac{N\,({\rm
 cm}^{-2})}{b\,({\rm km\ s}^{-1})}, 
\end{equation}

\noindent
and a Doppler parameter $b=30$ km s$^{-1}$ characteristic of the
\Lya\ forest absorptions at redshift $z \sim 2$ \citep{kim01}. 

This technique is efficient for data with $S/N \ga 1/f_{\rm thres} =
10$, so we applied the selection only to the LP QSO sample and  
to the sample of simulated spectra for which, in this case, we adopted
a Gaussian noise giving $S/N=50$.     

Fig.~\ref{fig:comp_autov_f01} shows the resulting correlation
functions. The amplitude of the observed $\xi^f_{\pa}$ has reduced to
about one third of the initial value but still shows a significant
clustering signal. This is an indication that matter is still
clustered also far from the most over-dense regions. The 
threshold  on $\tau_{\rm H\,I}$, transformed with eq.~(2), corresponds to 
$\rho/ \bar{\rho} \la 6.5$.

What is worth noting is that the amplitude of the simulated 
correlation function has decreased of about one half and is now fully  
compatible with the observed correlation function. This suggests that
strong lines are deficient in the simulated spectra so that their
contribution to the correlation function is smaller than for the
observed spectra. When they are excluded from the computation the two
correlation functions become consistent.


\subsection{Selection effects in the observed spectra}

There is the possibility that our observed sample is affected by a 
selection bias since the QSOs were required to be at close angular
separations and, in most cases, they have similar emission redshifts. 
As a consequence, the observed \loss\ could be biased toward more 
clustered regions than average.

We tried to estimate this effect on the auto correlation function by
excluding larger and larger portions of the observed spectra from the
QSO emissions (up to 20000 \kms). 
No significant decrease of the amplitude of the correlation function
has been detected.  

This test could prove inadequate as far as the sestet is concerned  
since the redshifts of some member QSOs are separated by up to $\sim
50000$ \kms\ (see Fig.~\ref{fig:sestet}).  
As a consequence, our measure could be affected by the transverse
proximity effect \citep[e.g.][]{jakobsen,wew} or by possible clustering
associated with the presence of a foreground QSO. Indeed, in the case
of Pair Q, we found a damped \Lya\ system in the higher redshift QSO
spectrum at the redshift of the other QSO \citep{vale02}. 
From a preliminary analysis of the sestet, strong \Lya\ absorptions
($\log N($\HI$)\sim 15$) are observed in the spectra of S3, S5 and S6
at $\sim 600-750$ \kms\ from the emission redshift of the lower
redshift QSO, S1, and a strong
\Lya\ with associated \CIV\ absorption is present in the spectrum of
S2, the furthest away QSO, at $\sim 1000$ \kms\ from the redshift of S1.  
Considering the other foreground object, S4, \Lya\ absorptions within
200 \kms\ of its redshift are observed along the \los\ to S2
($\log N($\HI$)\sim 14$) and S3 ($\log N($\HI$)\sim 15$) without
associated metals detected.  
As already mentioned in the Introduction, we will carry out a detailed
study of the coincidences among multiple \loss\ in a future paper,
with data of improved S/N.

In the following, we will use the auto correlation function
determined from the LP sample to avoid systematic uncertainties
introduced by the above mentioned possible selection effect.   


\section{Clustering of the flux across the line of sight}

In this section we exploit the potentialities of our 
sample of QSO pairs by determining the clustering properties of the
IGM across the \loss.  
The great advantage with respect to the correlation function along the
\los, in particular for a sample like ours showing a large variety of
pair separations, is that we have the guarantee of sampling true
spatial separations between the pixels, the effect of peculiar
velocities being negligible or absent.  
    
\subsection{The cross correlation function in redshift space}

As a first approach, we computed the cross correlation function
extending in a natural way the procedure adopted for the auto
correlation function. 

Every pixel along the \los\ is considered as an element of the density
field at the QSO angular position in the sky and at a distance from
the observer (comoving along the \los) corresponding to the 
redshift of the pixel:  

\begin{equation}
r_{\pa}(z) = \frac{c}{H_0} \int_0^z \frac{{\rm d}z'}{E(z')},  
\end{equation}

\noindent
where $E(z)$ is defined as:

\begin{equation}
E(z)  = \sqrt{\Omega_{0\rm{m}}\,(1+z)^3 + \Omega_{0\Lambda}}. 
\end{equation} 

In eq.~(5), there is the implicit hypothesis that peculiar velocities
contribute negligibly to the measured redshift in the \Lya\ forest. 
This statement is supported by recent measurements using QSO pairs    
\citep{rauch05}, and we will see {\it a posteriori} that the
comparison of the auto and cross correlation functions makes this
procedure allowable.

The cross correlation function of the transmitted flux between two
\loss\ at angular separation $\Delta\theta$ is defined as:

\begin{equation}
\xi^f_{\times}(\Delta r) =
\langle\delta_f(\theta,r_{\pa,1})
\delta_f(\theta+\Delta\theta,r_{\pa,2})\rangle,  
\end{equation}

\noindent
where, $\Delta r = \sqrt(r_{\pa,1}^2 + r_{\pa,2}^2 - 2\,
r_{\pa,1}\,r_{\pa,2}\,\cos\Delta\theta)$ is the spatial separation
between pixel 1 at $r_{\pa,1}$ along one \los\ and pixel 2 at
$r_{\pa,2}$ along the paired \los.

The error on the correlation function cannot be estimated
with a bootstrap technique, as done before, due to the limited number
of pairs contributing to each separation (in particular at the smaller
separations).   
In order to give an evaluation of the significance of the observed
signal, the same cross correlation function was computed replacing one
QSO in each pair with a control QSO, then repeating the operation
replacing the other QSO of every pair with another control QSO.  
The control sample is formed by the LP QSO spectra. We chose LP QSOs
at redshifts close to those of the original objects and shifted the
spectra in order to match it exactly.  
We derived an indicative error as the r.m.s. deviation from zero of
the two control functions in two regions:  $\Delta r < 5\
h^{-1}$ Mpc, for which the uncertainty is larger, and  $\Delta r
> 5\ h^{-1}$ Mpc. 

\begin{figure}
\includegraphics[height=8truecm,width=8truecm]{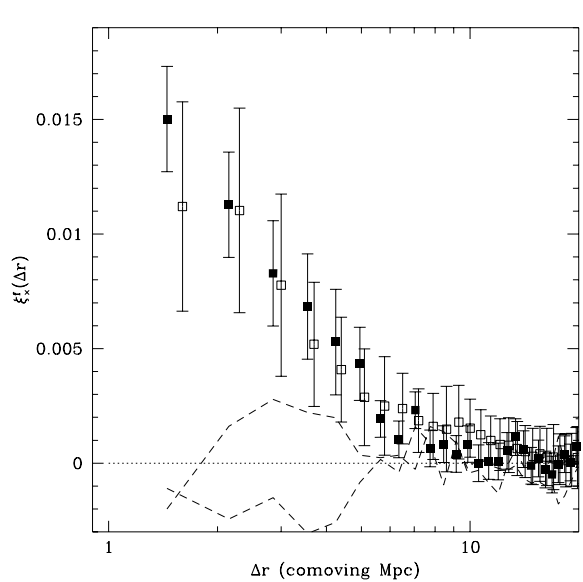}
  \caption{Cross correlation function of the transmitted flux for our
    sample of pairs and groups of QSOs (solid squares) 
    compared with the analogous correlation function computed with the
    sample of mock spectra (empty squares). The error bars on the
    simulated data are 1~$\sigma$ standard deviations for 50
    simulated samples while the error bars on the observed data have
    been obtained from the r.m.s. deviation of the two control samples
    plotted as dashed lines (see text). Simulated data are slightly
    shifted in $\Delta r$ for clarity.}\label{fig:comp_crosscf} 
\end{figure}

We computed the cross correlation function also for the sample of
mock spectra. The simulated spectra are characterised by the redshift 
of the output box and a velocity extent (see Section~2.2). 
In order to assign a redshift value to every pixel, we gave the
central pixel of every  spectrum the redshift of the corresponding
output box, then we numbered the pixels one by one transforming the
velocity size of the pixel into a redshift size. 
Once the redshifts were determined pixel by pixel, we followed the
same procedure adopted for the observed spectra for the 50 simulated
samples and computed the average cross correlation function and its
$1\ \sigma$ standard deviation.  

The result of our computation is shown in Fig.~\ref{fig:comp_crosscf}
compared with the observed cross correlation function of the pairs and
of the control sample. 
There is a very good agreement between the two functions. Only in the
first bin there is an indication of a lower value for the simulated
cross correlation with respect to the observed one.  
This is consistent with the result obtained for the auto correlation
function if we take into account the fact that the first bin of
$\xi^f_{\times}$ corresponds to the second bin of $\xi^f_{\pa}$. 



\subsection{The cross correlation coefficient}

A measure of the transverse clustering properties of the IGM which is
less affected by peculiar velocities is the flux cross correlation
coefficient,

\begin{equation}
\chi^f_{\times}(\Delta\theta)  = \langle \delta_f(\theta, v_{\pa}) 
\delta_f(\theta+\Delta\theta,v_{\pa}) \rangle, 
\label{ccc}
\end{equation}
 
\noindent
where, every pixel along one \los\ is correlated with the one
face-to-face in redshift space along the paired \los\ and the result
is averaged over all the pixels in the common redshift interval.  

Every pair of QSOs at angular separation $\Delta\theta$ gives one
value of $\chi^f_{\times}(\Delta\theta)$, and a sample with several
pairs at different separations, as is our sample, gives an estimate of the
correlation function.  
At a given redshift, the angular separation $\Delta\theta$ corresponds
to a velocity separation $\Delta v_{\perp}=c\,F(z)\,\Delta\theta$,
where  $c$ denotes the speed of light, and $F(z)$ is a dimensionless
function of redshift that includes all the dependence on the global
cosmological metric. 
In the cosmological model that we have adopted: 

\begin{equation}
F(z)  =  \frac{E(z) \int_0^z [{\rm d}z/E(z)]}{(1+z)}, 
\end{equation}

\noindent
and $E(z)$ is defined in eq.~(6).

\begin{figure}
\includegraphics[height=8truecm,width=8truecm]{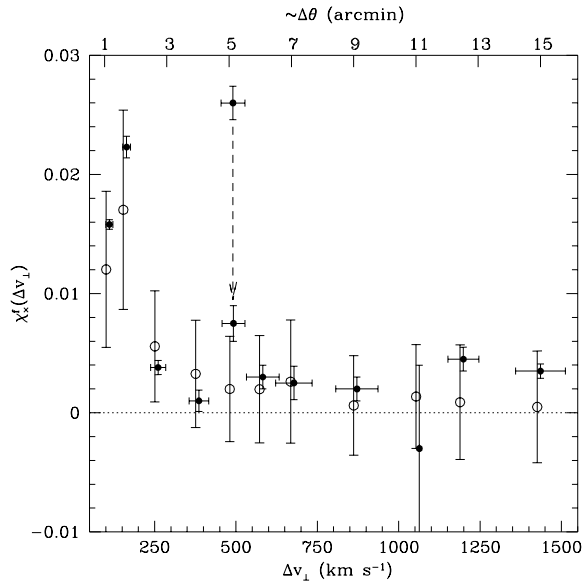}
  \caption{Comparison of the cross correlation coefficients 
  for our sample of observed spectra (solid dots) and of
  simulated ones (empty dots) as a function of the velocity
  separation, $\Delta v_{\pe} = c\,F(z)\,\Delta\theta$,    
  corresponding to the angular separation, $\Delta\theta$, of the QSO
  pairs. The angular separation computed at $z=1.8$ is reported in the
  top axes. 
  Observed values are shifted by 10 \kms\ for clarity. 
  The lower observed value at $\Delta v_{\pe} \simeq 500$ \kms\ was
  obtained excluding pair S1S3 and removing a strong coincident \Lya\
  line in the T1T3 pair (see text).  
  Error bars on the observed values along the x axis represent the
  velocity range covered by the considered pairs. Errors on the
  $\chi^f_{\times}$ represent the uncertainty on that measure for the observed
  values and the cosmic variance for the simulated values (see text).
  }\label{fig:comp_ccc} 
\end{figure}

We computed the range of velocity separations covered by each of our
pair of spectra then we grouped the pairs in velocity bins of variable
width and computed the average cross correlation coefficient for every
group. Given the small number of pairs in every group (a maximum of 3
QSO pairs) the uncertainties reported on these determinations are 
computed by applying a simple error propagation to eq.~(8), thus they
account only for the pixel statistics and the noise associated with
the transmitted flux but they are not representative of the true error
due to the cosmic variance.  

In the case of the simulated spectra, we had 50 realisations of each of
our QSO pairs so we could obtain in every velocity interval defined
for the observed pairs an average cross correlation coefficient with
its error, that in this case is the standard deviation of the
distribution of values. 

Results are shown in Fig.~\ref{fig:comp_ccc}. 
The two points at the smallest separations are given by Pair U + S5S6 
and Pair A, respectively. 
The two samples of values are in very good agreement except for the
bin at $\Delta v_{\perp} \sim 500$ \kms\ ($\simeq 5$ arcmin at $z\sim
1.8$) for which the observed value is $\sim 7\ \sigma$ above the
simulated value.  
The observed  $\chi^f_{\times}(\Delta\theta)$ in this velocity bin is
obtained by averaging two pairs: S1S3 and T1T3 for which
$\chi^f_{\times} \simeq 0.040\pm0.003$ and $0.0177\pm0.0015$,
respectively.  

\begin{figure}
\includegraphics[height=8truecm,width=8truecm]{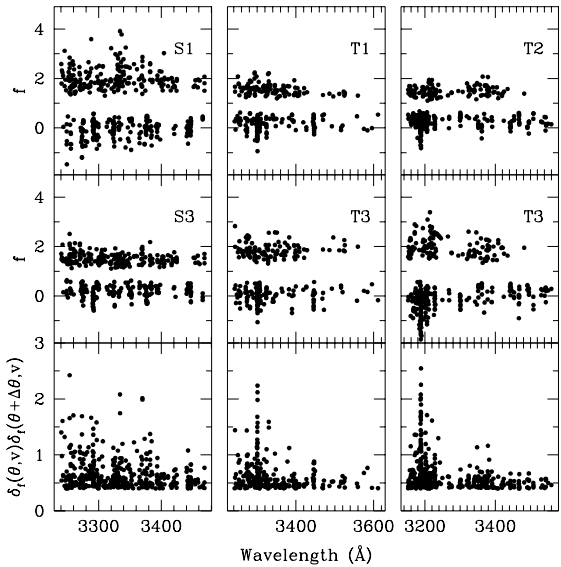}
  \caption{Selection of the pixels giving values $\delta_f(\theta, v_{\pa}) 
\delta_f(\theta+\Delta\theta,v_{\pa}) > 0.4$ for
    the pairs S1S3 (left), T1T3 (centre) and T2T3(right). The two
    upper plots show the values of the flux for the selected pixels in
    the indicated QSO while the lower plot report the value of the
    above product.} 
  \label{fig:ccc_5am}
\end{figure} 

The two pairs of spectra were searched for peculiar features that
could boost the signal by selecting those pixels for which
$\delta_f(\theta, v_{\pa}) \delta_f(\theta+\Delta\theta,v_{\pa}) >
0.4$. In Fig.~\ref{fig:ccc_5am} we show the results of this research.  
In the case of S1 (up left plot) about 44 per cent of selected pixels have
values $f>2$ or $f<0$, that is they are strongly affected by
noise (in the case of  S3 only $\sim 14$ per cent of pixels are in those range of 
values). As a consequence, the large value of the cross correlation
coefficient for pair S1S3 could be due to the low S/N in the spectrum
of S1 and cannot be determined reliably. 
On the other hand, in the case of pair T1T3, Fig.~\ref{fig:ccc_5am}
(centre plots) shows a feature 
giving a strong signal at $\lambda \simeq 3300$ \AA. We identified in
the spectra of T1 and T3 two coincident strong \HI\ \Lya\ absorptions
($\log N($\HI$) \sim 15$) at this wavelength, the former with a clear
associated \CIV\ doublet while for the second there is the possible
detection of the \CIV $\lambda\,1548$ line only. No corresponding \HI\
\Lya\ line is observed at this redshift along the third \los\ of the
triplet, T2. 
If we mask the redshift interval covered by the coincident lines and
recompute $\chi^f_{\times}$ for T1T3 we get: $0.0075\pm0.0015$, which is in
better agreement with the simulated value as shown in
Fig.~\ref{fig:comp_ccc}.  

In the right hand plots of Fig.~\ref{fig:ccc_5am}, we show
the case of pair T2T3 (at 8.9\arcmin\ angular separation) where a
coincident absorption system is present with characteristics very
similar to the one in pair T1T3, and the cross correlation coefficient
has a value consistent with zero. 
This suggests that the presence of a coincident absorption system
could be a necessary but not sufficient condition to explain a large 
value  of $\chi^f_{\times}$.   
%
It is interesting to note that also the recently published two point 
correlation function of the \CIV\ absorptions in the LP QSO spectra  
\citep{evan06} shows an  excess of clustering signal at $\sim 500$ 
\kms\ whose origin has not been satisfactorily explained.  



New observations are going to be carried out with UVES to
increase the S/N in the two QSO pair
spectra S1S3 and T1T3, in order to confirm the large value of the 
cross correlation
coefficient and possibly detect other coincident \CIV\ systems. 
On the other hand, we will try and collect other QSO pairs at the same
velocity separation to increase the statistics and we will look into
physical mechanisms that could explain our results.

\begin{figure}
\includegraphics[height=8truecm,width=8truecm]{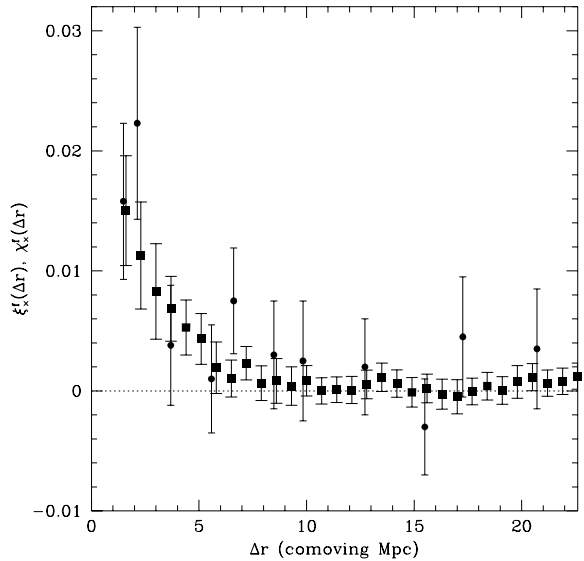}
  \caption{Cross correlation function of the transmitted flux,
    $\xi^f_{\times}$, for our sample of pairs and groups of QSOs
    (squares, slightly shifted in $\Delta r$) compared with the
    observed cross correlation 
    coefficients, $\chi^f_{\times}$ (dots), as a function of spatial
    separation (see text). Error bars both on $\chi^f_{\times}$ and
    $\xi^f_{\times}$ have been 
    determined from simulations. }\label{fig:spcf} 
\end{figure}

In Fig.~\ref{fig:spcf}, we compare the cross correlation coefficients
with the cross correlation function.  
The angular separation $\Delta \theta$ between two QSO \loss\ was
transformed into a comoving spatial separation, $\Delta r$, with the
formula:  

\begin{equation}
\Delta r = \frac{c\Delta \theta}{H_0} \int_0^z \frac{{\rm d}z'}{E(z')},  
\end{equation}

\noindent
where $E(z)$ was defined in eq.~(6). 
Considering the error bars computed from the simulated pairs, there is
a good agreement within 1~$\sigma$.    
The large variations from one data point to the other in the cross
correlation coefficients should be due mainly to the small number of
QSO pairs (between 1 and 3) contributing to each point. On the other
hand, the smoothness of $\xi^f_{\times}$ is artificially increased by
the fact that the values in the different bins are not independent.


\section{Discussion}

From the cosmological point of view, one of the most interesting and
challenging applications of the kind of calculations carried out in
this paper is the constraint on the geometry of the Universe (in
particular, the estimate of $\Omega_{0\Lambda}\ h^{-2}$) by  linking
angular separations and redshift differences in the hypothesis that the
observed correlation properties are isotropic
\citep[][]{ap,mcdonald_miralda,hui}. 

A large number of QSO pairs at different angular separations
\citep[see][]{mcdonald03} is needed to obtain a measure of the
cosmological parameters as accurate as those recently produced by the
{\it WMAP} team \citep{spergel}. This is why we do not attempt to
derive a measure of $\Omega_{0\Lambda}$ with the present sample. 

\begin{figure}
\includegraphics[height=8truecm,width=8truecm]{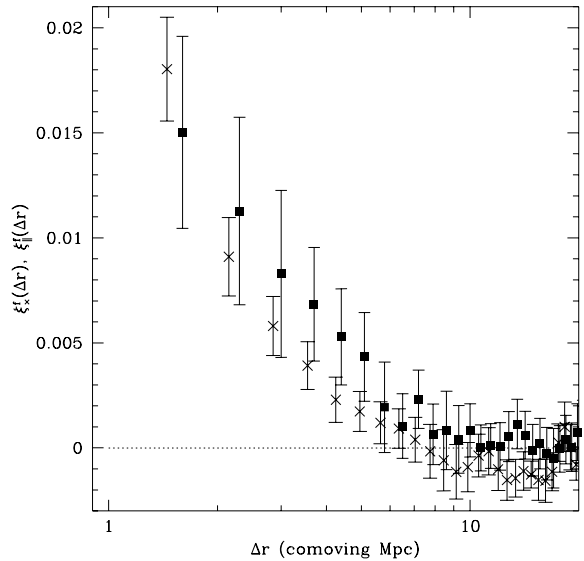}
  \caption{Comparison of the cross correlation function for
  our sample of QSO pairs (squares) with the auto correlation
  function computed for the LP QSO sample (crosses) as a
  function of comoving spatial separation across and along the \los,
  respectively. The cross correlation function is slightly shifted in
  $\Delta r$ for clarity}\label{fig:comp_auto_cross} 
\end{figure}

Nevertheless, a comparison of the observed auto and cross correlation
functions is interesting to qualitatively evaluate the distortion due
to peculiar velocities and the correctness of the adopted cosmological
parameters. 
The two functions are shown in Fig.~\ref{fig:comp_auto_cross}.  
Note that, due to the minimum angular separation between our QSO
pairs, there is no transverse clustering signal corresponding to the
first bin of the auto correlation (centred at $\Delta v=50$ \kms\ or
$\Delta r \simeq 740$ kpc). 
The correspondence between  $\xi^f_{\pa}$ and $\xi^f_{\times}$ is very
good suggesting that the adopted cosmological parameters are
reasonable and that peculiar velocities do not play a major role in
the IGM gas on velocity scales $\ga 100$ \kms.     

In the next section, we survey the behaviour of the simulated
correlation functions when varying three parameters
characterising the physical state of the gas. 
    
\subsection{Exploring the parameter space of the simulations}

\subsubsection{Effective optical depth}

As already explained in Section~2.2, the effective optical depth of a
simulated spectrum can be rescaled to a new value simply by increasing
or decreasing the optical depth of every pixel in the spectrum by a
constant factor, which is equivalent to increase or decrease the
column density of the absorption lines in the spectrum.  

\begin{table}
\centering
\caption{Effective optical depth adopted to build the three samples of
mock spectra used to compute the correlation functions in
Figs~\ref{fig:auto_taueff} and \ref{fig:cross_taueff}}
\label{taueff}
\begin{tabular}{@{}lcccc@{}}
\hline
QSO & z & $\tau_{\rm min}$ & $\tau_{\rm obs}$ & $\tau_{\rm max}$ \\
\hline
Sestet & 1.8 & 0.08 &  0.12  & 0.17 \\
Triplet & 1.8 & 0.08 &  0.10  & 0.17 \\
Pair U  & 1.8 & 0.08 &  0.10  & 0.17 \\
Pair A & 2.4 & 0.17 & 0.21 & 0.29 \\
Pair Q & 2.4 & 0.17 & 0.21 & 0.29 \\
\hline
\end{tabular}
\end{table}

Considering previous observations \citep{viel2004a} and the relations
computed by \citet{schaye03} for the redshift evolution of the
effective optical depth, we derived $3\,\sigma$ upper and lower limits
for $\tau_{\rm eff}$ at the redshifts of the simulated spectra, which
are reported in Table~\ref{taueff}.     
Then, we built two new samples of mock spectra starting from the
original spectra produced from the simulation, one with the minimum and
one with the maximum values of $\tau_{\rm eff}$ and recomputed the
auto and cross correlation functions. 
Each sample is formed by 50 realisations of the observed sample of
QSO pairs and we determined the average value for the correlation
function. 
  
\begin{figure}
\includegraphics[height=8truecm,width=8truecm]{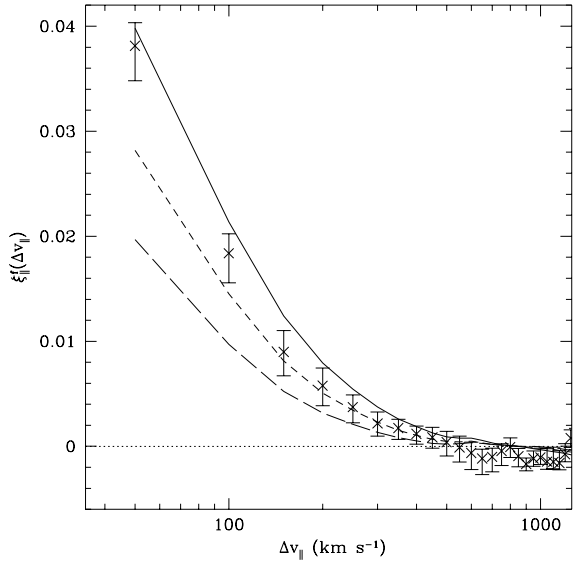}
  \caption{Auto correlation function for the LP sample (crosses)
  related to the average auto correlation functions of 
  three samples of simulated spectra with different $\tau_{\rm
  eff}$. The long dashed line traces the correlation function for the
  sample with the minimum optical depths, the dashed line is for the
  sample with the observed optical depths and the solid line is for
  the sample with the maximum optical depths (see Table~\ref{taueff}). 
}\label{fig:auto_taueff} 
\end{figure}

\begin{figure}
\includegraphics[height=8truecm,width=8truecm]{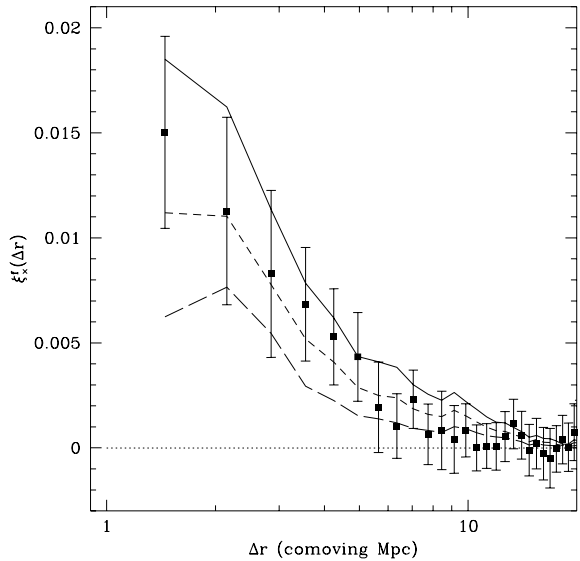}
  \caption{Cross correlation function for our sample of QSOs (solid
  squares) compared with the average cross correlation functions of
  three samples of simulated spectra with different $\tau_{\rm
  eff}$. The long dashed line traces the correlation function for the
  sample with the minimum optical depths, the dashed line is for the
  sample with the observed optical depths and the solid line is for
  the sample with the maximum optical depths (see Table~\ref{taueff}) 
}\label{fig:cross_taueff} 
\end{figure}

Our results are shown in Figs~\ref{fig:auto_taueff} and
\ref{fig:cross_taueff}.  At the smallest scale, probed only by the
auto correlation function, the agreement between observed and
simulated clustering signal improves adopting the larger effective
optical depths.  
However, at the Jeans scale, corresponding to the first bin of
$\xi^f_{\times}$ and to the bin centred at $100$
\kms\ of $\xi^f_{\pa}$,  the larger $\tau_{\rm eff}$ possibly
overestimates the clustering amplitude. 

Since the effective optical depth measured in the original simulation
is in very good agreement with the one observationally determined from
QSO \Lya\ forests (see Secion.~2.2), we infer that the discrepancies
between simulations and observations are not due to a wrong estimate
of $\tau_{\rm eff}$ but, as already mentioned in Section~4, to the
deficiency of strong lines in the simulations. 
In the spectra with the larger $\tau_{\rm eff}$ the column
density of all the lines is increased and the average flux is
decreased, causing a general increase of the amplitude of the
correlation function, partially compensating the small number of
strong lines.   

In order to verify our statement, a quantitative analysis of the
difference in the \Lya\ column density distribution function (the
number of absorption lines per unit column density per unit redshift)
of observed and simulated spectra  \citep[as done by][ for small
simulated boxes]{theuns02} is necessary. Furthermore, more pairs at
small separations ($<1$\arcmin) would allow to decrease the error bars
and test if the behaviour of  clustering is the same as along the \los.

\begin{figure}
\includegraphics[height=8truecm,width=8truecm]{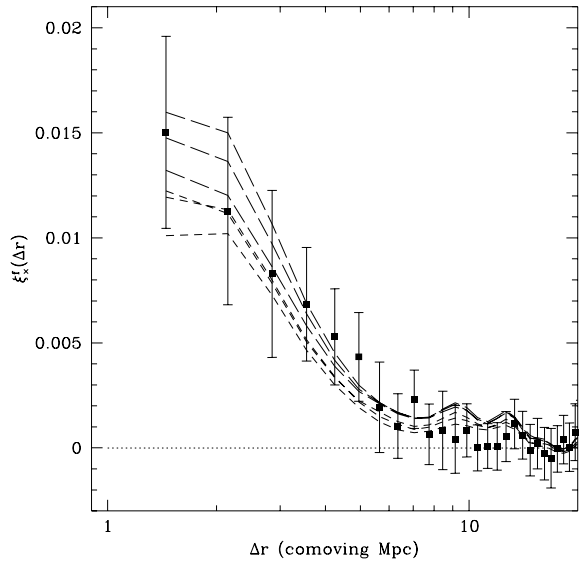}
  \caption{Cross correlation function for our sample of pairs (solid
  squares)  compared with the average cross correlation functions of 
  samples of simulated spectra with different temperature and exponent
  $\gamma$ of the temperature-density relation (eq.~(1)). Short dashed
  lines are for simulations in which $\gamma=1.6$ and temperature
  is $T=8000$, 22000 and 50000 K going from the upper to the lower  
  curve. The long dashed lines represents simulations with $\gamma=1.1$
  while the temperatures are the same as before but they increase from
  the lower to the upper curve   
}\label{fig:cross_Tgamma} 
\end{figure}

\subsubsection{Temperature and $\gamma$}
  
The influence on the flux cross correlation function of the
temperature and of the power-law index of the temperature-density relation
adopted for the gas (see eq.~(1)) are explored. 
The aim of this analysis is to point out the modifications of the
simulated correlation function when the parameters are varied, and not 
to derive precise constraints on the same parameters from the
comparison with observations. With a running time of 2-3 weeks for
simulation, it is impossible to run an extensive parameter study for
different thermal histories, thereby we decided to rescale {\sl a
  posteriori} the temperature-density relation at a given output by
assuming that all the gas particles obey this relation. 
We then recomputed the ionization fractions for each gas particle with
the proper UV background. 
This approximation does not account for the effects that
the corresponding change in the gas pressure would have on the gas
distribution. However we explicitly checked that the differences in the
correlation function for the fiducial run and for a rescaled temperature
relation with the same $T-\gamma$ are of the order of 10 per cent 
\citep[see][ for the effect of the rescaling on the flux
  power spectrum]{viel2004a}. 

We built six sets of simulated spectra each one formed by 50
realisations of the observed sample of pairs. Every set is
characterised by a value of 
$\gamma$ (1.1 or 1.6) and a value of the temperature (8000, 22000, and
50000 K), while the effective optical depth is the one used for the
main sample (refer to Section~2.2). 

In Fig.~\ref{fig:cross_Tgamma}, we plot together the observed
$\xi^f_{\times}$ with the average correlation
functions computed from the six groups of simulated spectra. 
At the smallest separation probed by the observations, the predictions
of the different models are well separated and could be tested with a
larger sample of close QSO pairs. 
The models with $\gamma=1.1$ give a larger clustering signal than
those with $\gamma=1.6$ for all temperatures and are in better
agreement with the data at the small scales. 
An improved concordance between observations and simulations for
lower values of $\gamma$ was observed also in the case of the flux
probability distribution function estimated from the \Lya\ forest
(James Bolton, private communication). 

The exponent of the temperature-density relation is determined by the
ionization history of the gas. \citet{huignedin} demonstrated that
instant reionizations of \HI\ occurring at redshifts decreasing from
$z= 10$ to 5 would imply shallower and shallower $\gamma$ at redshifts
$z\sim 3$. However, also for very late reionizations $\gamma$ should
have increased above values $\sim 1.4$ by $z\sim 2$. 
\citet*{ricotti00} studied the evolution with redshift of the equation
of state of the gas and showed that a reionization of \HeII\ occurring 
at $z\sim 3$ would cause a sudden increase in the gas temperature and a
corresponding decrease in the value of $\gamma$ (their fig.~13). 
Observationally, this has been studied mainly by using the
distribution of Doppler parameters and column densities of the \Lya\
forest lines \citep[e.g.][]{schaye00,kim02}, but also with metal
absorptions  \citep[e.g.][]{songaila98,vladilo}.    
Our result supports previous observational evidences of a second
reheating of the Universe happening at $z\sim 3$, most likely due to
the reionization of \HeII\ and it requires confirmation by a larger
sample of observed spectra and a set of more refined mock spectra.

\section{Conclusions} 

In this paper, we exploited the capabilities of the largest sample
of high resolution UVES spectra of QSO pairs to study the
3-dimensional distribution properties of baryonic matter in the IGM as
traced by the transmitted flux in the QSO \HI\ \Lya\ forests. 
Our sample is formed by 21 QSO pairs evenly distributed between
angular separations of $\sim 1$ and 14 arcmin, with \Lya\ forests at
a median redshift $z \simeq 1.8$. 
We selected also 8 UVES QSO spectra from the ESO Large Program `The
Cosmic Evolution of the IGM' \citep{bergeron04} to compute the
correlation function along the line of sight and to be used as a
control sample for the cross correlation function (see Section~5.1).   
We compared the observed sample with a set of mock spectra drawn 
from a cosmological hydro-simulation run in a box of $120\ h^{-1}$
comoving Mpc, adopting the cosmological parameters of the concordance
model.  
The simulated sample reproduces 50 different realisations of the
observed sample. 
 
In the following, we resume our main results: 
\par\noindent 
1. The clustering properties of matter in the IGM are the same in the 
direction parallel and transverse to the line of sight when using the 
parameters of the concordance cosmology to transform the angular
distance into velocity separation. As an implication, peculiar
velocities in the absorbing gas are likely smaller than $\sim 100$
\kms.    
\par\noindent
2. Matter in the IGM is clustered on scales smaller than $\sim 200$
\kms\ or about $3\ h^{-1}$ comoving Mpc. We verified that the clustering
signal is significant also for the slightly over-dense gas ($\tau_{\rm
  H\,I} \ga 2.3$ or $\rho/ \bar{\rho} \la 6.5$) although on smaller
scales.   
\par\noindent 
3. The simulated correlation functions are consistent with the
observed analogous quantities at the $3\,\sigma$ level, although they
systematically predict lower clustering at the smaller scales.  The
agreement becomes better when only the lower density regions are
selected for the computation or when the effective optical depth of
the simulated spectra is fixed to a larger value (marginally
consistent with previous extensive observational results on the
redshift evolution of the effective optical depth).  These are
indications of a deficiency of strong absorption lines in the
simulated spectra that needs further investigation.  
\par\noindent 
4. We observed an improved consistency between observations and
simulations also when a lower $\gamma$ is adopted in the
equation of state of the gas, $T=T_0\, (\rho/\bar{\rho})^{\gamma-1}$
($\gamma=1.1$ instead of the standard $1.6$). This result hints to a 
late \HeII\ reionization epoch whose effects on the IGM could still be 
measured at the redshifts investigated here.
\par\noindent
5. We measured an enhanced clustering signal for
the cross correlation coefficient
at a transverse velocity separation $\Delta v_{\pe} \sim 500$ \kms. This
velocity distance matches the line split of the \CIV\ doublet. 
We propose to gather more observational and theoretical material
to shed light on this result.  

From the analysis carried out in the present study, we evidenced the 
need of increasing our sample of observed QSO pairs, in particular at
the small angular separations ($\sim1$ arcmin), but also at $\Delta
v_{\pe} \simeq 500$ \kms\ in order to confirm the overdensity of
coincident \Lya\ lines at this transverse distance. 
On the other hand, we realised the necessity of a dedicated effort on
the comparison between observed and simulated QSO spectra, in
particular as far as the statistics of absorption lines is concerned
(number density, Doppler parameter and column density distribution,
etc.).

\section*{Acknowledgments}
We are indebted to an anonymous referee whose comments were extremely
helpful to improve the clarity of this paper. 
We are grateful to Pierluigi Monaco for stimulating comments and a
critical reading of the manuscript.  
It is our pleasure to thank Evan Scannapieco and Emmanuel Rollinde for
interesting discussions.
SL was partly supported by the Chilean {\sl Centro de Astrof\'\i sica}
FONDAP No. 15010003, and by FONDECYT grant N$^{\rm o} 1030491$. 
The simulations were run on the COSMOS (SGI Altix 3700) supercomputer
at the Department of Applied Mathematics and Theoretical Physics in
Cambridge.  COSMOS is a UK-CCC facility which is supported by HEFCE
and PPARC.

\bibliographystyle{aa} 
\bibliography{aamnem99,myref} 


\label{lastpage}

\end{document}